\RequirePackage{fix-cm}
\documentclass[twocolumn,epjc3]{svjour3}
\smartqed  
\RequirePackage{graphicx}
\RequirePackage{mathptmx}      
\usepackage{amsmath, amssymb}
\usepackage{cite}
\journalname{Eur. Phys. J. C}
\begin{document}

\title{A study of the measurement precision of the Higgs boson decaying into tau pairs at the ILC
}


\author{Shin-ichi Kawada\thanksref{e1,addr1}
\and
Keisuke Fujii\thanksref{addr2}
\and
Taikan Suehara\thanksref{addr3}
\and
Tohru Takahashi\thanksref{addr1}
\and
Tomohiko Tanabe\thanksref{addr4}
}

\thankstext{e1}{e-mail: s-kawada@huhep.org}


\institute{Advanced Sciences of Matter (AdSM), Hiroshima University, 1-3-1, Kagamiyama, Higashi-Hiroshima, Hiroshima, 739-8530, Japan \label{addr1}
\and
High Energy Accelerator Research Organization (KEK), 1-1, Oho, Tsukuba, Ibaraki, 305-0801, Japan \label{addr2}
\and
Graduate School of Science, Kyushu University, 6-10-1, Hakozaki, Higashi-ku, Fukuoka, 812-8581, Japan \label{addr3}
\and
International Center for Elementary Particle Physics (ICEPP), The University of Tokyo, 7-3-1, Hongo, Bunkyo-ku, Tokyo, 113-0033, Japan \label{addr4}
}

\date{Received: XXX / Accepted: XXX}

\maketitle

\begin{abstract}
We evaluate the measurement precision of the production cross section times the branching ratio of the Higgs boson decaying into tau lepton pairs at the International Linear Collider (ILC).
We analyze various final states associated with the main production mechanisms of the Higgs boson,
the Higgs-strahlung and $WW$-fusion processes.
The statistical precision of the production cross section times the branching ratio is estimated to be
2.6{\%} and 6.9{\%} for the Higgs-strahlung and $WW$-fusion processes, respectively,
with the nominal integrated luminosities assumed in the ILC Technical Design Report;
the precision improves to 1.0{\%} and 3.4{\%} with the running scenario including possible luminosity upgrades.
The study provides a reference performance of the ILC for future phenomenological analyses.
\end{abstract}

\section{Introduction}

After the discovery of the Higgs boson by the ATLAS and the CMS experiments at the LHC~\cite{ATLAS, CMS}, the investigation of the properties of the Higgs boson has become an important target of study in particle physics.
In the Standard Model (SM), the coupling of the Higgs boson to the matter fermions, \textit{i.e.}, the Yukawa couplings, is proportional to the fermion mass.
The Yukawa couplings can deviate from the SM prediction in the presence of new physics beyond the SM.
Recent studies indicate that the deviations from the SM could be at the few-percent level if there is new physics at the scale of around 1 TeV~\cite{deviate}.
It is therefore desired to measure the Higgs couplings as precisely as possible in order to probe new physics.

In this study, we focus on Higgs boson decays into tau lepton pairs ($h \to \tau ^+ \tau ^-$) at the International Linear Collider (ILC).
This decay has been studied by the ATLAS and the CMS experiments, who reported a combined signal yield consistent with the SM expectation, with a combined observed significance at the level of $5.5 \sigma$~\cite{HTTevidence1, HTTevidence2, HTTevidence3}.
The purpose of this study is to estimate the projected ILC capabilities of measuring the $h \to \tau ^+ \tau ^-$ decay mode in final states resulting from the main Higgs boson production mechanisms in $e^+ e^-$ collisions.
Existing studies on $h \to \tau ^+ \tau ^-$ decays at $e^+ e^-$ collisions~\cite{previous1, previous2} did not take into account some of the relevant background processes or were based on a Higgs boson mass hypothesis which differs from the observed value, both of which are addressed in this study.
We assume the ILC capabilities for the accelerator and the detector as documented in the ILC Technical Design Report (TDR)~\cite{TDR1, TDR2, TDR3_1, TDR3_2, TDR4} together with its running scenario published recently~\cite{PhysicsCase, BeamPolOpe}.
The results presented in this paper will be useful for future phenomenological studies.

The contents of this paper are organized as follows.
In Section 2, we describe the ILC and the ILD detector concept, and the analysis setup.
The event reconstruction and selection at center-of-mass energies of 250 GeV and 500 GeV are discussed in Sections 3 and 4.
Section 5 describes the prospects for improving the measurement precision with various ILC running scenarios.
We summarize our results in Section 6.

\section{Analysis conditions}

\subsection{International Linear Collider}

The ILC is a next-generation electron--positron linear collider.
It covers a center-of-mass energy ($\sqrt{s}$) in the range of 250--500 GeV and can be extended to $\sqrt{s}=1$~TeV.
In the ILC design, both the electron and positron beams can be polarized, which allow precise measurements of the properties of the electroweak interaction. 
The details of the machine design are summarized in the ILC Technical Design Report (TDR)~\cite{TDR1, TDR2, TDR3_1, TDR3_2, TDR4}.

The ILC aims to explore physics beyond the SM via precise measurements of the Higgs boson and the top quark
as well as to search for new particles within its energy reach.
The center-of-mass energies and integrated luminosities which are foreseen are summarized in Table~\ref{MainParameter}.
The numbers for the nominal running scenario are taken from the ILC TDR~\cite{TDR1}.
The numbers for scenarios including energy and luminosity upgrades are based on studies in Refs.~\cite{PhysicsCase, BeamPolOpe}.

\begin{table}[tb]
\centering
\caption{Typical integrated luminosities $L$ and center-of-mass energies $\sqrt{s}$ of the ILC~\cite{TDR1, PhysicsCase, BeamPolOpe}.}
\begin{tabular}{ccc}\hline
Scenario & $\sqrt{s}$ (GeV) & $L$ (fb$^{-1}$) \\ \hline
Nominal & 250 & 250 \\
 & 500 & 500 \\
\hline 
Luminosity upgrade & 250 & 2000 \\
 & 500 & 4000 \\
\hline
\end{tabular}
\label{MainParameter}
\end{table}

\subsection{Production and decay of the Higgs boson}

\begin{figure}[tb]
\centering
\includegraphics[width = 4.2truecm]{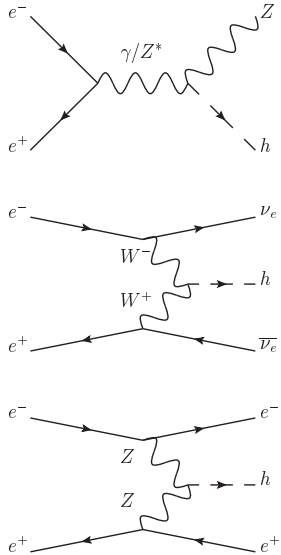}
\caption{Diagrams of the main production mechanisms of the Higgs boson in $e^+ e^-$ collisions.
Top: Higgs-strahlung ($Zh$) process. Middle: $WW$-fusion process. Bottom: $ZZ$-fusion process.}
\label{HiggsProduction}
\end{figure}

\begin{figure}[tb]
\centering
\includegraphics[width = 7.5truecm]{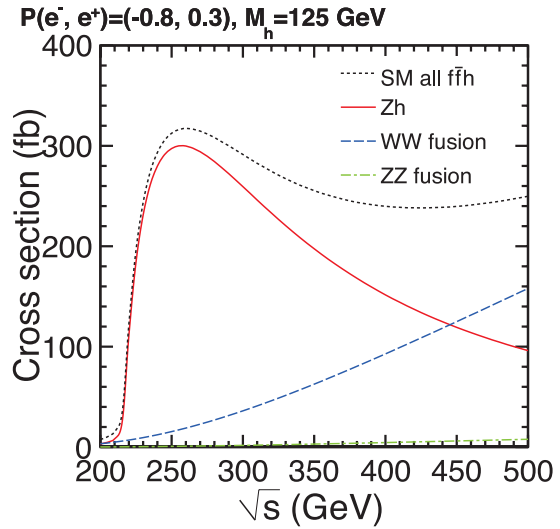}
\caption{Cross sections of the Higgs boson production as a function of $\sqrt{s}$ in the electron--positron interaction.}
\label{HiggsXsec}
\end{figure}

Figure~\ref{HiggsProduction} shows the diagrams for the main production mechanisms of the Higgs boson in $e^+ e^-$ collisions.
The cross sections of Higgs boson production calculated by \verb|WHIZARD|~\cite{Whizard} with a Higgs mass of 125 GeV are shown in Figure~\ref{HiggsXsec}, where polarizations of $-80{\%}$ and $+30{\%}$ for the electron and positron beams are assumed, and initial state radiation is taken into account.

For the calculation of the production cross section and the subsequent decay of the signal processes of $e^+ e^- \to f\overline{f}h \to f\overline{f}\tau ^+ \tau ^-$, where $f$ denotes a fermion, we use an event generator based on \verb|GRACE|~\cite{GRACE1, GRACE2}.
The effect of beamstrahlung is implemented according to the calculation by \verb|GuineaPig|~\cite{GuineaPig}, which simulates $e^+ e^-$ beam--beam interactions, with the beam parameters described in the TDR~\cite{GDE}.
Initial state radiation is incorporated following the prescription developed by the ILC Event Generator Working Group~\cite{TDR4, ISR}.
To handle the spin correlation of tau pairs from the Higgs boson decay, \verb|GRACE| is interfaced with \verb|TAUOLA|~\cite{Yokoyama, Tauola1, Tauola2, Tauola3}.
The decays of other short-lived particles and the hadronization of quarks and gluons are handled by \verb|PYTHIA|~\cite{Pythia}.

\subsection{Background processes}

For background processes, we use common Monte-Carlo (MC) samples for SM processes previously prepared for the studies presented in the ILC TDR~\cite{TDR4}.
The event samples include $e^+ e^- \to 2f$, $e^+ e^- \to 4f$, $e^+ e^- \to 6f$, and $e^+ e^- \to f\overline{f}h$.
The event generation of these processes is performed with \verb|WHIZARD|~\cite{Whizard}, in which beamstrahlung, initial state radiation, decay of short-lived particles, and hadronization are taken into account in the same way as described in the previous section for the signal process.
The background processes from $\gamma \gamma$ interactions with hadronic final states, in which photons are produced by beam--beam interactions, are generated on the basis of the cross section model in Ref.~\cite{overlay}.
We find that the interactions between electron or positron beams and beamstrahlung photons, \textit{i.e.}, $e^{\pm} \gamma \to e^{\pm} \gamma$, $e^{\pm} \gamma \to 3f$, and $e^{\pm} \gamma \to 5f$,
have negligible contributions to background.

\subsection{Detector Model}

The detector model used in this analysis is the International Large Detector (ILD), which is one of the two detector concepts described in the ILC TDR.
It is a general-purpose $4\pi$ detector designed for particle flow analysis\footnote{The particle flow algorithm aims at achieving the best attainable jet energy resolution by making one-to-one matching of charged particle tracks with calorimetric clusters so as to restrict the use of calorimetric information, which is in general less precise than tracker information, to neutral particles.
This requires highly granular calorimeters and a tracking system with high performance pattern recognition for events with high particle multiplicity.}, aiming at best possible jet energy resolution.

The ILD model consists of layers of sub-detectors surrounding the interaction point.
One finds, from the innermost to the outer layers, a vertex detector (VTX), a silicon inner tracker (SIT), a time projection chamber (TPC), a silicon envelope tracker (SET), an electromagnetic calorimeter (ECAL), and a hadron calorimeter (HCAL), all of which are put inside a solenoidal magnet providing a magnetic field of 3.5~T.
The return yoke of the solenoidal magnet has a built-in muon system.
The ILD design has not yet been finalized.
In this analysis, we assume the following configurations and performance.
The VTX consists of three double layers of silicon pixel detectors with radii at 1.6~cm, 3.7~cm and 6~cm.
Each silicon pixel layer provides a point resolution of 2.8~$\mu$m.
The TPC provides up to 224 points per track over a tracking volume with inner and outer radii of 0.33~m and 1.8~m.
The SIT and SET are used to improve the track momentum resolution by adding precise position measurements just inside and outside of TPC.
The ECAL consists of layers of tungsten absorbers interleaved with silicon layers segmented into $5 \times 5$ mm$^2$ cells, has an inner radius of 1.8 m, and has a total thickness of 20 cm corresponding to 24 radiation length. 
The HCAL consists of layers of steel absorbers interleaved with scintillator layers segmented into $3 \times 3$ cm$^2$ cells and has an outer radius of 3.4 m corresponding to 6 interaction length.
Additional silicon trackers and calorimeters are located in the forward region to assure hermetic coverage down to 5~mrad from the beam line.
The key detector performance of the ILD model is summarized in Table~\ref{ILDPerformance}.
Details of the ILD model and the particle flow algorithm are found in Refs~\cite{TDR4, PFA}.

\begin{table}[tb]
\centering
\caption{Summary of the performance of the ILD detector model.}
\begin{tabular}{cc}\hline
Name & Value \\ \hline
Impact parameter resolution & $5 \oplus \dfrac{10}{p\sin ^{3/2}\theta} \mu$m \\
Momentum resolution & $2 \times 10^{-5} \oplus \dfrac{1 \times 10^{-3}}{P_t\sin \theta}$ GeV/$c$ \\
Jet energy resolution & $\sim \dfrac{30}{\sqrt{E (\mathrm{GeV})}}${\%} \\
\hline
\end{tabular}
\label{ILDPerformance}
\end{table}

\subsection{Detector simulation and event reconstruction}

In this study, we assume a Higgs boson mass of 125 GeV,
a branching ratio of the Higgs boson decay into tau pairs ($\mathrm{BR}(h \to \tau ^+ \tau ^-)$) of $6.32{\%}$~\cite{NNLO}, and beam polarizations of $-80{\%}$ and $+30{\%}$ for the electron and the positron beams, respectively.

We perform a detector simulation with \verb|Mokka|~\cite{Mokka}, a \verb|Geant4|-based~\cite{Geant4} full detector simulator, with the ILD model for all signal and background processes,
with the exception of the $e^+ e^- \to e^+ e^- + 2f$ process at $\sqrt{s} = 500$ GeV,
for which \verb|SGV| fast simulation~\cite{SGV} is used.
The event reconstruction and physics analysis are performed within the \verb|MARLIN| software framework~\cite{Marlin}, in which events are reconstructed using track finding and fitting algorithms, followed by a particle flow analysis using the \verb|PandoraPFA| package~\cite{PFA}.

\section{Analysis at the center-of-mass energy of 250 GeV}

At $\sqrt{s} =$ 250 GeV, the Higgs-strahlung ($e^+ e^- \to Zh$) process dominates the SM Higgs production, as shown in Figure~\ref{HiggsXsec}.
The $WW$-fusion and $ZZ$-fusion cross sections are negligible at this energy.
We take into account $e^+ e^- \to f\overline{f}h$ (excluding the $h \to \tau ^+ \tau ^-$ signal),
$e^+ e^- \to 2f$, and $e^+ e^- \to 4f$ for the background estimation.
The $\gamma \gamma \to$ hadrons background is overlaid onto the MC samples with an average of 0.4 events per bunch crossing~\cite{overlay}.
An integrated luminosity of 250 fb$^{-1}$ is assumed for the results in this section.

There are four main signal modes:
$e^+ e^- \to q\overline{q}h$,
$e^+ e^- \to e^+ e^- h$,
$e^+ e^- \to \mu ^+ \mu ^- h$,
and $e^+ e^- \to \nu \overline{\nu}h$.
For our $\sqrt{s}=$ 250~GeV results,
we report on the first three of these modes.
We do not quote the results for the $\nu \overline{\nu} h$ mode
as we find that it suffers from background processes with neutrinos in the final state.
We do not analyze the $e^+ e^- \to \tau^+ \tau^- h$ mode in this study.


\subsection{$e^+ e^- \to q\overline{q}h$}

\textbf{Reconstruction of isolated tau leptons and the $Z \to q\overline{q}$ decay}

For the $q\overline{q} h$ mode,
we first identify the tau leptons using a dedicated algorithm developed for this topology.
%
%
The algorithm proceeds as follows.
\begin{enumerate}
\item The charged particle with the highest energy is chosen as a working tau candidate.
\item The tau candidate is combined with the most energetic particle (charged or neutral) satisfying the following two conditions: the angle $\theta_i$ between the particle and the tau candidate satisfies $\cos\theta_i>0.99$; and
the combined mass, calculated from the sum of the four momenta of the particle and the tau candidate, does not exceed 2 GeV.  The four momentum of this particle is then added to that of the tau candidate.
\item Step 2 is repeated until there are no more particles left to combine.
      The resulting tau candidate is then set aside.
\item The algorithm is repeated from Step 1 until there are no more charged particles left.
\end{enumerate}
%
A tau candidate is accepted if
the number of charged particles with track energy greater than 2~GeV is equal to one or three,
the net charge is equal to $\pm 1$, and the total energy is greater than 3~GeV.
Furthermore, an isolation requirement is applied as follows.
A cone of half-angle $\theta_c$, with $\cos\theta_c=0.95$, is defined around the direction of the tau momentum.
The tau candidate is accepted if the energy sum of all particles inside the cone (excluding those forming the tau candidate) does not exceed 10\% of the tau candidate energy.
%
%
%
%
We require exactly two final tau candidates with opposite charges.
This results in a selection efficiency of 49.3{\%} for the $q\overline{q} \tau ^+ \tau ^-$ signal events.

\begin{figure}[tb]
\centering
\includegraphics[width = 7.5truecm]{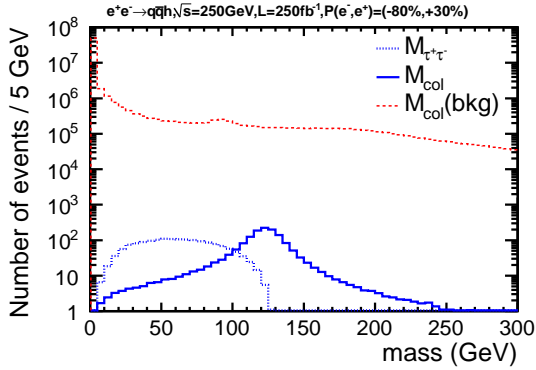}
\caption{Distributions of the invariant mass of the reconstructed tau lepton pairs at $\sqrt{s}=250$~GeV
for the $e^+e^-\to q\overline{q} h$ mode.
$M_{\tau ^+ \tau ^-}$ and $M_{\mathrm{col}}$ stand for the tau pair masses
before and after the collinear approximation, respectively, for the signal.
$M_{\mathrm{col}} (\mathrm{bkg})$ is the tau pair mass with the collinear approximation for the background.
}
\label{taupairmass_qqh250}
\end{figure}

After the tau candidates are identified, the neutrino energy is recovered by using the collinear approximation~\cite{colapp}.
Because tau leptons from a Higgs boson decay are highly boosted, it is reasonable to assume that the tau momentum and the neutrino momentum are nearly parallel.
Under this assumption, the energy of the two neutrinos, one from each tau decay, can be solved by requiring that the overall transverse momentum of the event is balanced in two orthogonal directions.
The neutrino reconstructed in this way is added to the tau candidate.
%
Figure~\ref{taupairmass_qqh250} shows the invariant mass distributions of the tau pairs without ($M_{\tau ^+ \tau ^-}$) and with ($M_{\mathrm{col}}$) the collinear approximation for the events containing two tau lepton candidates with opposite charges.
With the collinear approximation, a clear peak is visible at 125 GeV for signal events.
The $M_{\mathrm{col}}$ distribution for background events with the same criteria is also shown.

The Durham jet clustering algorithm~\cite{Durham} is applied to the remaining particles to reconstruct the two jets from the $Z$ boson decay.

\noindent \textbf{Event selection}

We perform a pre-selection over the reconstructed events, followed by a multivariate analysis.
The pre-selection is designed to reduce background while keeping most of the signal.
The events are pre-selected according to the following criteria.
The $Z\to q\overline{q}$ candidate and the $h\to \tau^+\tau^-$ candidate are successfully reconstructed.
The total number of charged particles is at least 9.
The visible energy of the event, $E_{\mathrm{vis}}$, lies in the range of 105 GeV $< E_{\mathrm{vis}} <$ 255 GeV.
The visible mass of the event, $M_{\mathrm{vis}}$, is greater than 95 GeV.
The sum of the magnitude of the transverse momentum of all visible particles, $P_{t,\mathrm{sum}}$, is greater than 40 GeV.
The thrust of the event is less than 0.97.
The $Z$ candidate dijet has an energy, $E_{Z}$, in the range of 60 GeV $< E_{Z} <$ 175 GeV
and has an invariant mass, $M_{Z}$, in the range of 35 GeV $<M_{Z}<$ 160 GeV.
The angle between the two jets, $\theta_{jj}$, satisfies $\cos\theta_{jj}<0.5$.
The recoil mass against the $Z$ boson, computed as
$M_{\mathrm{recoil}} = \sqrt{ (\sqrt{s}-E_{Z})^2-|\vec{p}_{Z}|^2 }$,
is in the range of 65 GeV $< M_{\mathrm{recoil}} <$ 185 GeV.
The Higgs candidate tau pair before the collinear approximation
has an energy, $E_{\tau^+\tau^-}$, less than 140 GeV
and an invariant mass, $M_{\tau^+\tau^-}$, in the range of 5 GeV $<M_{\tau^+\tau^-}<$ 125 GeV.
The angle between the two tau candidates, $\theta_{\tau^+\tau^-}$, satisfies $\cos\theta_{\tau^+\tau^-}<-0.1$.
The tau pair after the collinear approximation has
an energy, $E_{\mathrm{col}}$, in the range of 30 GeV $< E_{\mathrm{col}} <$ 270 GeV
and an invariant mass, $M_{\mathrm{col}}$, in the range of 15 GeV $< M_{\mathrm{col}} <$ 240 GeV.


We use a multivariate analysis using Boosted Decision Trees (BDTs) as implemented in the Toolkit for Multivariate Data Analysis~\cite{TMVA} of the ROOT framework~\cite{ROOT}.
The input variables are
\begin{itemize}
\item $E_{\mathrm{vis}}$, $P_{t,\mathrm{vis}}$, $\cos \theta _{\mathrm{miss}}$, where
$P_{t,\mathrm{vis}}$ is the magnitude of the visible transverse momentum and
$\theta _{\mathrm{miss}}$ is the angle of the missing momentum with respect to the beam axis;
\item $M_{Z}$, $\cos \theta _{jj}$,
$M_{\mathrm{recoil}}$,
$\cos \theta _Z$, where $\theta _Z$ is the angle of the $Z$ candidate momentum with respect to the beam axis;
\item $M_{\tau ^+ \tau ^-}$, $E_{\tau ^+ \tau ^-}$, $\cos \theta _{\tau ^+ \tau ^-}$, $\cos \theta _{\mathrm{acop}}$, where $\theta _{\mathrm{acop}}$ is the acoplanarity angle between the two tau candidates;
\item $\displaystyle \sum _{\tau ^+ , \tau ^-} \log _{10}|d_0 / \sigma_{d_0}|$,
      $\displaystyle \sum _{\tau ^+ , \tau ^-} \log _{10}|z_0 / \sigma_{z_0}|$,
where $d_0/\sigma_{d_0}$ and $z_0/\sigma_{z_0}$ are respectively the transverse and longitudinal impact parameters 
of the most energetic track in the tau candidate
divided by their respective uncertainty estimated from the track fit;
\item $M_{\mathrm{col}}$, $E_{\mathrm{col}}$, and $\cos \theta _{\mathrm{col}}$, where $\theta _{\mathrm{col}}$ is the angle of the Higgs candidate momentum with the collinear approximation measured from the beam axis.
\end{itemize}
\begin{figure}[tb]
\centering
\includegraphics[width = 7.5truecm]{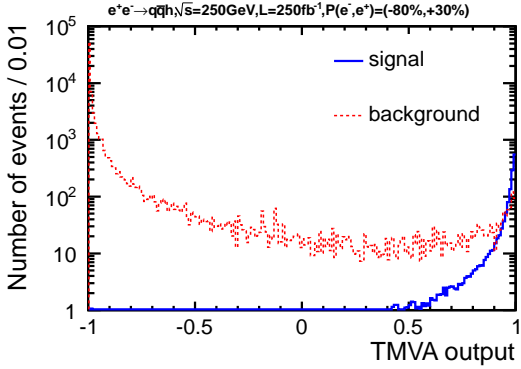}
\caption{Distributions of the multivariate discriminant from training Boosted Decision Trees
for the $e^+ e^- \to q\overline{q}h$ mode, shown for the signal and the total background.
}
\label{qqh250_BDTGoutput}
\end{figure}
The BDTs are trained using a set of statistically independent signal and background samples.
The distribution of the resulting multivariate discriminant is shown in Figure~\ref{qqh250_BDTGoutput}.
We apply a final selection on the multivariate discriminant that maximizes the signal significance defined as $S/\sqrt{S+B}$, where $S$ and $B$ are the number of signal and background events, respectively.
The final selected sample consists of 1232 signal and 543 background events.
The estimated event yields before and after the selection are summarized in Table~\ref{CutTable_qqh250}.
The signal selection efficiency is 37{\%} with a signal significance of 29, which corresponds to a statistical precision of $\Delta (\sigma \times \mathrm{BR}) / (\sigma \times \mathrm{BR}) = 3.4{\%}$.

\begin{table}[tb]
\centering
\caption{Event yields estimated for the $e^+ e^- \to q\overline{q}h$ mode at $\sqrt{s} = 250$ GeV, assuming an integrated luminosity of 250 fb$^{-1}$ and beam polarizations of $P(e^- , e^+) = (-0.8, +0.3)$,
shown for the signal and the background processes.
The signal contribution ($h\to\tau^+\tau^-$) is removed from the $f \overline{f} h$ process.
``No cut'' is the number of events corresponding to the production cross section times the integrated luminosity.
``Pre-selected'' is the number of events after the pre-selection for the multivariate analysis.
``Final'' is the number of events after the selection on the multivariate discriminant.
}
\begin{tabular}{ccccc}\hline
 & Signal & $f\overline{f}h$ & $2f$ & $4f$ \\ \hline
No cut & 3318 & $7.649 \times 10^4$ & $2.863 \times 10^7$ & $1.736 \times 10^8$ \\
Pre-selected & 1451 & 3526 & 2316 & $6.940 \times 10^4$ \\
Final & 1232 & 22.0 & 9.3 & 512.0 \\
\hline
\end{tabular}
\label{CutTable_qqh250}
\end{table}

\subsection{$e^+ e^- \to e^+ e^- h$}

\textbf{$Z$ boson and tau lepton reconstruction}

For the $e^+ e^- h$ mode,
we first reconstruct the $e^+e^-$ pair that forms a $Z$ boson candidate.
A reconstructed particle is identified as an electron or a positron if its track momentum ($P_{\mathrm{trk}}$) and
its associated energy deposits in the ECAL ($E_{\mathrm{ECAL}}$) and HCAL ($E_{\mathrm{HCAL}}$)
satisfy the following criteria:
\begin{align*}
E_{\mathrm{ECAL}} / ( E_{\mathrm{ECAL}} + E_{\mathrm{HCAL}} ) &> 0.96, \\
( E_{\mathrm{ECAL}} + E_{\mathrm{HCAL}} ) / P_{\mathrm{trk}} &> 0.6.
\end{align*}
For the particles that are identified as electrons or positrons,
we further require that $|d_0 / \sigma_{d_0}| < 6$ and $|z_0 / \sigma_{z_0}| < 3$,
to reduce the electrons from secondary decays such as the tau lepton decays from the Higgs boson.
We also require the track energy to be greater than 10 GeV, which removes the contamination from the $\gamma \gamma \to$ hadron background.
The $e^+e-$ pair whose combined mass is closest to the $Z$ boson mass is selected as the $Z$ boson candidate.
To improve the mass and energy resolutions, the momenta of nearby neutral particles are added to that of the $Z$ candidate if their angle $\theta$ measured from at least one of the $e^\pm$ satisfies $\cos \theta > 0.999$.
The fraction of $e^+ e^- \tau ^+ \tau ^-$ signal events that survive the $Z$ boson selection is 61{\%}.

We apply a tau finding algorithm to the remaining particles.
Compared with the $q\overline{q} h$ mode, the algorithm is simpler due to the absence of hadronic jet activities aside from the tau decays.
Starting with the charged particle with the highest energy as a working tau candidate,
we define a cone around its momentum vector with a half-angle of $\theta_c=1.0$ rad.
Particles inside the cone are combined with the tau candidate if the combined mass remains smaller than 2 GeV.
The tau candidate is then set aside, and the tau finding is repeated until there are no more charged particles left.
The tau candidates are then separated into two categories according to its charge.
Within each category, the tau candidate with the highest energy is selected.
The chosen $\tau^+ \tau^-$ pair forms the Higgs candidate.
Finally, the collinear approximation is applied to the selected tau candidates. 

\noindent \textbf{Event selection}

A pre-selection is applied with the following requirements before proceeding with the multivariate analysis.
The $Z\to e^+e^-$ candidate and the $h\to \tau^+\tau^-$ candidate are successfully reconstructed.
The total number of charged tracks is 8 or fewer, which ensures statistical independence from the $q\overline{q}h$ mode.
The visible energy is in the range of 100 GeV $< E_{\mathrm{vis}} <$ 280 GeV.
The visible mass is in the range of 85 GeV $< M_{\mathrm{vis}} <$ 275 GeV.
The sum of the magnitude of the transverse momentum of all visible particles, $P_{t,\mathrm{sum}}$, is greater than 35 GeV.
The $Z\to e^+e^-$ candidate has an energy in the range of 40 GeV $< E_{Z} <$ 160 GeV
and an invariant mass in the range of 10 GeV $< M_{Z} < 145$ GeV.
The recoil mass against the $Z$ boson, $M_{\mathrm{recoil}}$, is greater than 50 GeV.


We then apply a multivariate analysis using BDTs using the following input variables:
\begin{itemize}
\item $M_{\mathrm{vis}}$, $E_{\mathrm{vis}}$, $\cos \theta _{\mathrm{miss}}$, $\cos \theta _{\mathrm{thrust}}$, where $\theta _{\mathrm{thrust}}$ is the angle of the thrust axis with respect to the beam axis;
\item $M_{Z}$, $M_{\mathrm{recoil}}$;
\item $M_{\tau ^+ \tau ^-}$, $\cos \theta _{\tau ^+ \tau ^-}$, $\cos \theta _{\mathrm{acop}}$;
\item $\displaystyle \sum _{\tau ^+ , \tau ^-} \log _{10}|d_0 / \sigma_{d_0}|$,
  and $\displaystyle \sum _{\tau ^+ , \tau ^-} \log _{10}|z_0 / \sigma_{z_0}|$.
\end{itemize}
A final selection on the multivariate discriminant is applied to maximize the signal significance, giving 76.3 signal and 44 background events.
The final signal selection efficiency is 44{\%}.
The estimated event yields before and after the selection are summarized in Table~\ref{CutTable_eeh250}.
The signal significance is estimated to be 7.0, corresponding to a statistical precision of
$\Delta (\sigma \times \mathrm{BR}) / (\sigma \times \mathrm{BR}) = 14.4{\%}$.

\begin{table}[tb]
\centering
\caption{Event yields estimated for the $e^+ e^- \to e^+ e^- h$ mode at $\sqrt{s} = 250$ GeV, assuming an integrated luminosity of 250 fb$^{-1}$ and beam polarizations of $P(e^- , e^+) = (-0.8, +0.3)$.
Refer to Table~\ref{CutTable_qqh250} for the row definitions.
}
\begin{tabular}{ccccc}\hline
 & Signal & $f\overline{f}h$ & $2f$ & $4f$ \\ \hline
No cut & 175.1 & $7.964 \times 10^4$ & $2.863 \times 10^7$ & $1.736 \times 10^8$ \\
Pre-selected & 109.4 & 60.2 & $3.334 \times 10^4$ & $1.169 \times 10^4$ \\
Final & 76.3 & 4.2 & 0 & 39.9 \\
\hline
\end{tabular}
\label{CutTable_eeh250}
\end{table}

\subsection{$e^+ e^- \to \mu ^+ \mu ^- h$}

\textbf{$Z$ boson and tau lepton reconstruction}

The reconstruction procedure of this mode is similar to that of the $e^+ e^- h$ mode,
with the electron identification replaced by the muon identification.
The muons are identified by requiring
\begin{align*}
E_{\mathrm{ECAL}} / ( E_{\mathrm{ECAL}} + E_{\mathrm{HCAL}} ) < 0.5, \\
( E_{\mathrm{ECAL}} + E_{\mathrm{HCAL}} ) / P_{\mathrm{trk}} < 0.6. \notag
\end{align*}
We additionally require the identified muons to satisfy
$|d_0 / \sigma_{d_0}| < 3$ and $|z_0 / \sigma_{z_0}| < 3$,
and to have a track energy greater than 20 GeV.
The efficiency for selecting such muon pairs in
$\mu^+\mu^- \tau^+\tau^-$ signal events is 92{\%}.
The tau lepton reconstruction is the same as in the $e^+ e^- h$ mode.

\noindent \textbf{Event selection}

The following pre-selection requirements are applied before proceeding with the multivariate analysis.
The $Z\to \mu^+\mu^-$ candidate and the $h\to \tau^+\tau^-$ are successfully reconstructed.
The total number of charged tracks is 8 or fewer.
The visible energy is in the range of 105 GeV $< E_{\mathrm{vis}} <$ 280 GeV.
The visible mass is in the range of 85 GeV $< M_{\mathrm{vis}} <$ 275 GeV.
The sum of the magnitude of the transverse momentum of all visible particles, $P_{t,\mathrm{sum}}$, is greater than 35 GeV.
The $Z\to \mu^+\mu^-$ candidate has an energy in the range of 45 GeV $< E_{Z} <$ 145 GeV
and an invariant mass in the range of 25 GeV $< M_{Z} <$ 125 GeV.
The recoil mass against the $Z$ boson, $M_{\mathrm{recoil}}$, is greater than 75 GeV.
The invariant mass of the tau pair system before the collinear approximation, $M_{\tau ^+ \tau ^-}$, is smaller than 170 GeV.


A multivariate analysis with BDTs is applied to the pre-selected events using the following input variables:
\begin{itemize}
\item $M_{\mathrm{vis}}$, $E_{\mathrm{vis}}$, $P_{t,\mathrm{vis}}$;
\item $M_{Z}$, $\cos \theta _Z$, $M_{\mathrm{recoil}}$;
\item $M_{\tau ^+ \tau ^-}$, $E_{\tau ^+ \tau ^-}$, $\cos \theta _{\tau ^+ \tau ^-}$, $M_{\mathrm{col}}$;
\item $\displaystyle \sum _{\tau ^+ , \tau ^-} \log _{10}|d_0 / \sigma_{d_0}|$,
  and $\displaystyle \sum _{\tau ^+ , \tau ^-} \log _{10}|z_0 / \sigma_{z_0}|$.
\end{itemize}
We apply a final selection on the multivariate discriminant to maximize the signal significance,
and obtain 101.9 signal and 31 background events.
The final signal selection efficiency is 62{\%}.
The estimated event yields before and after the event selection are shown in Table~\ref{CutTable_mmh250}.
The signal significance is estimated to be 8.8, corresponding to a statistical precision of
$\Delta (\sigma \times \mathrm{BR}) / (\sigma \times \mathrm{BR}) = 11.3{\%}$.

\begin{table}[tb]
\centering
\caption{Event yields estimated for the $e^+ e^- \to \mu ^+ \mu ^- h$ mode at $\sqrt{s} = 250$ GeV, assuming an integrated luminosity of 250 fb$^{-1}$ and beam polarizations of $P(e^- , e^+) = (-0.8, +0.3)$.
Refer to Table~\ref{CutTable_qqh250} for the row definitions.
}
\begin{tabular}{ccccc}\hline
 & Signal & $f\overline{f}h$ & $2f$ & $4f$ \\ \hline
No cut & 164.6 & $7.965 \times 10^4$ & $2.863 \times 10^7$ & $1.736 \times 10^8$ \\
Pre-selected & 132.8 & 63.5 & 4182 & 8011 \\
Final & 101.9 & 2.2 & 0 & 29.0 \\
\hline
\end{tabular}
\label{CutTable_mmh250}
\end{table}



\section{Analysis at the center-of-mass energy of 500 GeV}

At $\sqrt{s} = 500$ GeV, both the $WW$-fusion and the Higgs-strahlung processes have sizable contributions
to the total signal cross section.
We take into account the $e^+ e^- \to f\overline{f}h$ (except $h \to \tau ^+ \tau ^-$), $e^+ e^- \to 2f$, $e^+ e^- \to 4f$, and $e^+ e^- \to 6f$ processes as backgrounds.
The $\gamma \gamma \to$ hadron background is overlaid onto the signal and background MC samples, assuming an average rate of 1.7 events per bunch crossing~\cite{overlay}.
The analysis in this section assumes an integrated luminosity of 500 fb$^{-1}$.
We report our results on the $e^+ e^- \to q\overline{q}h$ and $e^+ e^- \to \nu \overline{\nu}h$ modes.
We do not give results for the $e^+ e^- \to e^+ e^- h$ and $e^+ e^- \to \mu ^+ \mu ^- h$ modes,
as they do not contribute significantly to the overall sensitivity due to their small cross sections.

\subsection{$e^+ e^- \to q\overline{q}h$}

\textbf{Reconstruction of isolated tau leptons and the $Z \to q\overline{q}$ decay}

\begin{figure}[tb]
\centering
\includegraphics[width = 7.5truecm]{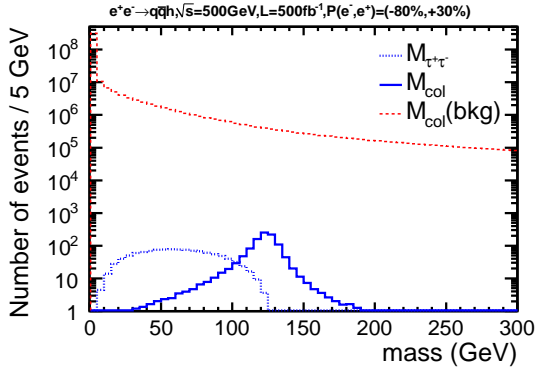}
\caption{Distributions of the invariant mass of the reconstructed tau lepton pairs at $\sqrt{s}=500$~GeV
for the $e^+e^- \to q\overline{q} h$ mode.
$M_{\tau ^+ \tau ^-}$ and $M_{\mathrm{col}}$ stand for the tau pair masses
before and after the collinear approximation, respectively,
for the signal process.
$M_{\mathrm{col}} (\mathrm{bkg})$ is the tau pair mass with the collinear approximation for the background processes.
}
\label{taupairmass_qqh500}
\end{figure}

We start with the tau finding, following the same procedure described in Section 3.1.
We additionally require the tau candidate to have an energy greater than 4~GeV.
The energy of the neutrino from tau decays is corrected using the collinear approximation as before,
resulting in a clear peak around the Higgs boson mass as can be seen in Figure~\ref{taupairmass_qqh500}.
We find that 54{\%} of $q\overline{q}\tau^+\tau^-$ signal events
survive the requirement of finding exactly one pair of $\tau^+\tau^-$.

The invariant mass of all the particles,
except those belonging to the two identified tau candidates,
should be consistent with the $Z$ boson mass;
however, a shift to a higher-mass value is observed,
due to the presence of non-negligible background particles from
$\gamma \gamma \to$ hadron events contaminating signal events.
In order to mitigate the effect of these background particles,
we use the $k_T$ clustering algorithm~\cite{kT1, kT2} implemented in the
\verb|FastJet| package~\cite{FastJet} with a generalized jet radius of $R = 0.9$.
The jets that are formed along the beam axis are then discarded.
The remaining particles are clustered into two jets by using the
Durham clustering algorithm to reconstruct the $Z$ boson decay.

\noindent \textbf{Event selection}

To facilitate the multivariate analysis, we impose the following pre-selections.
The $Z\to q\overline{q}$ candidate and the $h\to \tau^+\tau^-$ candidate are successfully reconstructed.
The total number of charged tracks is between 8 and 70.
The visible energy of the event is in the range of 140 GeV $< E_{\mathrm{vis}} <$ 580 GeV.
The visible mass of the event is in the range of 120 GeV $< M_{\mathrm{vis}} <$ 575 GeV.
The sum of the magnitude of the transverse momentum of all visible particles, $P_{t,\mathrm{sum}}$, is greater than 70 GeV.
The thrust of the event is less than 0.98.
The $Z$ candidate dijet has an energy in the range of 50 GeV $< E_{Z} <$ 380 GeV
and has an invariant mass in the range of 5 GeV $<M_{Z}<$ 350 GeV.
The recoil mass against the $Z$ boson is in the range of 40 GeV $< M_{\mathrm{recoil}} <$ 430 GeV.
The Higgs candidate tau pair before the collinear approximation
has an energy, $E_{\tau^+\tau^-}$, less than 270 GeV
and an invariant mass, $M_{\tau^+\tau^-}$, less than 180 GeV,
and the angle between the two tau candidates satisfies $\cos\theta_{\tau^+\tau^-}<0.7$.
The tau pair after the collinear approximation has
an energy in the range of 40 GeV $< E_{\mathrm{col}} <$ 430 GeV
and an invariant mass, $M_{\mathrm{col}}$, which is less than 280 GeV.
%

A multivariate analysis with BDTs is applied using the following input variables:
\begin{itemize}
\item $E_{\mathrm{vis}}$, $P_{t,\mathrm{sum}}$, $P_{\mathrm{vis}}$, where $P_{\mathrm{vis}}$ is the magnitude of the visible momentum;
\item $M_{Z}$, $E_{Z}$, $\cos \theta _{jj}$, $\cos \theta _Z$, $M_{\mathrm{recoil}}$;
\item $M_{\tau ^+ \tau ^-}$, $\cos \theta _{\tau ^+ \tau ^-}$, $M_{\mathrm{col}}$, $E_{\mathrm{col}}$;
\item $\displaystyle \sum _{\tau ^+ , \tau ^-} \log _{10} |d_0 / \sigma_{d_0}|$,
  and $\displaystyle \sum _{\tau ^+ , \tau ^-} \log _{10} |z_0 / \sigma_{z_0}|$.
\end{itemize}
After choosing the optimum threshold on the multivariate discriminant to maximize the signal significance,
we are left with 782 signal and 335 background events.
The final signal selection efficiency is 37{\%}.
The event yields before and after the selection are summarized in Table~\ref{CutTable_qqh500}.
The signal significance is found to be 23.4, corresponding to a statistical precision of
$\Delta (\sigma \times \mathrm{BR}) / (\sigma \times \mathrm{BR}) = 4.3{\%}$.

\begin{table*}[tb]
\centering
\caption{Event yields estimated for the $e^+ e^- \to q\overline{q}h$ mode at $\sqrt{s} = 500$ GeV,
assuming an integrated luminosity of 500 fb$^{-1}$ and beam polarizations of $P(e^- , e^+) = (-0.8, +0.3)$.
Refer to Table~\ref{CutTable_qqh250} for the row definitions.
}
\begin{tabular}{cccccc}\hline
 & Signal & $f\overline{f}h$ & $2f$ & $4f$ & $6f$ \\ \hline
No cut & 2131 & $1.266 \times 10^5$ & $1.320 \times 10^7$ & $9.989 \times 10^8$ & $6.929 \times 10^5$ \\
Pre-selected & 1088 & 2889 & $3.013 \times 10^4$ & $1.144 \times 10^5$ & $1.737 \times 10^4$ \\
Final & 782.1 & 17.6 & 1.5 & 275 & 41 \\
\hline
\end{tabular}
\label{CutTable_qqh500}
\end{table*}

\subsection{$e^+ e^- \to \nu \overline{\nu}h$}

\noindent \textbf{Tau pair reconstruction}

The tau finding algorithm proceeds in the same way as described for the $e^+ e^- h$ mode in Section 3.2,
except that the half-angle of the cone $\theta_c$ around the most energetic track is modified to 0.76 rad.
The most energetic positively and negatively charged tau candidates are combined to form a Higgs boson candidate.

\noindent \textbf{Event selection}


For the $\nu \overline{\nu} h$ mode,
it is necessary to suppress the large background
coming from the $e^+ e^- \to e^+ e^- + 2f$ processes.
We apply the following requirements to mitigate this background.
A tau lepton pair $\tau ^+ \tau ^-$ is successfully reconstructed.
The total number of tracks is less than 10.
There is at least one charged track with a transverse momentum greater than 3 GeV
and at least one charged track with an energy greater than 5 GeV.
The missing momentum angle with respect to the beam axis satisfies $|\cos\theta _{\mathrm{miss}}|<0.98$.
The acoplanarity angle between the two tau candidates satisfies $\cos\theta _{\mathrm{acop}}<0.98$.
At this point, 94{\%} of the $e^+ e^- \to e^+ e^- + 2f$ background is eliminated,
while retaining 85{\%} of the signal events.


The following additional pre-selections are applied before the multivariate analysis.
The visible energy is in the range of 10 GeV $< E_{\mathrm{vis}} <$ 265 GeV.
The visible mass is in the range of 5 GeV $< M_{\mathrm{vis}} <$ 235 GeV.
The missing mass, $M_{\mathrm{miss}}$, is greater than 135 GeV.
The sum of the magnitude of the transverse momentum of all visible particles, $P_{t,\mathrm{sum}}$, is greater than 10 GeV.
The Higgs candidate tau pair before the collinear approximation
has an energy, $E_{\tau^+\tau^-}$, less than 240 GeV
and an invariant mass, $M_{\tau^+\tau^-}$, of less than 130 GeV.
The angle between the two tau candidates satisfies $\cos\theta_{\tau^+\tau^-}<0.8$.
A requirement on the transverse impact parameter of the tau candidate
which gives a smaller value of the two is applied, such that
$\min |d_0 / \sigma_{d_0}| > 0.01$.


A multivariate analysis with BDTs is applied using the following input variables:
\begin{itemize}
\item Number of tracks with energy greater than 5 GeV;
\item Number of tracks with transverse momentum greater than 5 GeV;
\item $M_{\mathrm{vis}}$, $E_{\mathrm{vis}}$, $P_{t,\mathrm{vis}}$, $P_{t,\mathrm{sum}}$,
      $\cos \theta _{\mathrm{thrust}}$, $\cos \theta _{\mathrm{miss}}$;
\item $M_{\tau ^+ \tau ^-}$, $E_{\tau ^+ \tau ^-}$, $\cos \theta _{\tau ^+ \tau ^-}$, $\cos \theta _{\mathrm{acop}}$;
\item $\log _{10} \min | d_0 / \sigma_{d_0}| $.
\end{itemize}
We obtain 1642 signal and $1.11 \times 10^4$ background events after
optimizing the selection on the multivariate discriminant.
The final signal selection efficiency is 30{\%}.
The event yields before and after the selection are summarized in Table~\ref{CutTable_nnh500}.
The signal significance is 14.5, corresponding to a statistical precision of
$\Delta (\sigma \times \mathrm{BR}) / (\sigma \times \mathrm{BR}) = 6.9{\%}$.

\begin{table*}[tb]
\centering
\caption{Event yields estimated for the $e^+ e^- \to \nu \overline{\nu} h$ mode at $\sqrt{s} = 500$ GeV,
assuming an integrated luminosity of 500 fb$^{-1}$ and beam polarizations of $P(e^- , e^+) = (-0.8, +0.3)$.
Refer to Table~\ref{CutTable_qqh250} for the row definitions.
}
\begin{tabular}{cccccc}\hline
 & Signal & $f\overline{f}h$ & $2f$ & $4f$ & $6f$ \\ \hline
No cut & 5534 & $1.232 \times 10^5$ & $1.320 \times 10^7$ & $9.989 \times 10^8$ & $6.929 \times 10^5$ \\
Pre-selected & 3623 & 1543 & $5.957 \times 10^4$ & $1.756 \times 10^7$ & 990.8 \\
Final & 1642 & 65.5 & 379 & $1.043 \times 10^4$ & 238 \\
\hline
\end{tabular}
\label{CutTable_nnh500}
\end{table*}

\begin{figure}[tb]
\centering
\includegraphics[width = 7.5truecm]{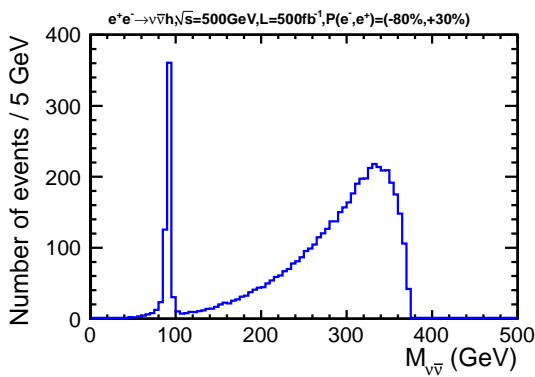}
\caption{Distribution of neutrino-pair invariant mass, $M_{\nu \overline{\nu}}$,
using event generator information
for the $e^+ e^- \to \nu \overline{\nu}h$ mode at $\sqrt{s}=500$~GeV.
The effect of detector simulation is not applied.}
\label{mc_Zmass_before_nnh500}
\end{figure}

In this mode, the $e^+ e^- \to Zh \to \nu \overline{\nu}h$ process
and the $e^+ e^- \to \nu \overline{\nu} h$ process via $WW$-fusion
are expected to be the dominant contributions.
The effect of the interference between these two processes is studied
using the distribution of the invariant mass of the neutrino pair
$M_{\nu \overline{\nu}}$ computed from event generator information
as shown in Figure~\ref{mc_Zmass_before_nnh500}.
A clear peak around the $Z$ boson mass is visible,
with a small contribution underneath it coming from the tail from higher masses,
indicating that the interference of the
$e^+ e^- \to Zh \to \nu \overline{\nu}h$ process and the $WW$-fusion is small.
We hence split the events into two categories based on this generator-level variable,
and define events with
$M_{\nu \overline{\nu}} < 120$ GeV as ``$e^+ e^- \to Zh$'' events,
and those with $M_{\nu \overline{\nu}} > 120$ GeV as ``$WW$-fusion'' events.
We find that the 1642 signal events after the final selection is composed of
13{\%} $e^+ e^- \to Zh$ events and 87{\%} $WW$-fusion events.
The selection efficiencies for $e^+ e^- \to Zh$ and $WW$-fusion events are
33.5{\%} and 29.2{\%}, respectively.

\section{Discussion}

\subsection{Precision with the ILC running scenarios}

We now discuss the prospects of the measurement precision
with the ILC running scenarios proposed in Refs.~\cite{BeamPolOpe, PhysicsCase}
by extrapolating the results presented in the previous sections.
Table~\ref{Actual_Int_Lumi} summarizes the integrated luminosities for various center-of-mass energies
and beam polarizations for three different scenarios we consider.
\begin{table*}[tb]
\centering
\caption{Integrated luminosity (in fb$^{-1}$) for various beam polarizations ($P(e^-)$, $P(e^+)$)
at $\sqrt{s} =$ 250 GeV and 500 GeV for three running scenarios.
``Nominal" is the scenario described in the TDR~\cite{TDR1} and used in the main results of this paper.
``Initial" and ``Full" are the scenarios proposed in Refs.~\cite{PhysicsCase, BeamPolOpe}.}
\begin{tabular}{c|c|cccc|c}\hline
Scenario & $\sqrt{s}$ (GeV) & ($-80 ,+30$) & ($+80, -30$) & ($-80, -30$) & ($+80, +30$) & Total \\ \hline
Nominal & 250 & 250 & 0 & 0 & 0 & 250 \\
 & 500 & 500 & 0 & 0 & 0 & 500 \\
\hline
Initial & 250 & 337.5 & 112.5 & 25 & 25 & 500 \\
 & 500 & 200 & 200 & 50 & 50 & 500 \\
\hline
Full & 250 & 1350 & 450 & 100 & 100 & 2000 \\
 & 500 & 1600 & 1600 & 400 & 400 & 4000 \\
\hline
\end{tabular}
\label{Actual_Int_Lumi}
\end{table*}

In order to estimate the statistical precision of the cross section times the branching ratio measurements
with electron and positron beam polarizations other than $(-0.8, +0.3)$ used in the previous sections,
we need to know the corresponding selection efficiencies for the signal and background processes.
In the following, we assume the same selection efficiencies obtained in the previous sections for
all of these beam polarizations, although in principle
the angular distributions of the final states may depend on the beam polarizations.
This assumption is nevertheless justified as follows.
The $e^+ e^- \to Zh$ process is mediated by the $s$-channel $Z$ boson exchange
with the vector or the axial vector coupling,
which forbids the same-sign helicity states $(\pm 1, \pm 1)$,
while giving more or less the same angular distributions for the
opposite-sign helicity states $(\mp 1, \pm 1)$.
On the other hand, the $WW$-fusion process proceeds only through the left-right helicity states $(-1, +1)$,
since the $W$ boson couples only to the left-handed $e^-$ and the right-handed $e^+$.
For the signal processes, therefore, their angular distributions stay the same for the
active (\textit{i.e.} opposite) helicity states,
independently of the choice of beam polarizations.
The same reasoning applies to the background processes with the
$s$-channel $\gamma /Z$ exchange or those involving $W$ bosons coupled to the initial state $e^+$ or $e^-$.
On the other hand, the processes involving $t$-channel photon exchange or photon-photon interactions
do not forbid the same-sign helicity states.
However, since the probability of finding an electron and a positron in the same-sign helicity states
is the same for both the $(-0.8, +0.3)$ and $(+0.8, -0.3)$ beam polarizations,
the efficiencies for such background processes with the same-sign helicity states should also be the same.
In our estimation, we do not use the results of the $(\pm 0.8, \pm 0.3)$ beam polarizations,
since the signal cross sections are small and the integrated luminosities
collected at these beam polarizations are foreseen to be small.
Under these assumptions, the selection efficiencies will not depend on the choice of beam polarizations.
%
We can then estimate the projected statistical precision for other scenarios by
calculating the number of signal and background events with
the production cross sections and the integrated luminosities for
individual beam polarizations, according to the running scenarios.
The result from this estimation is summarized in Table~\ref{SummaryTable}.

\subsection{Precision of the $h\to\tau^+\tau^-$ branching ratio}

So far, we discussed the precision of the production cross section times the branching ratio, which is the primary information we will obtain from the experiments.
Here, we discuss the prospects for measuring the branching ratio itself.
At the ILC, the production cross section for the Higgs-strahlung process can be separately measured using the recoil mass technique~\cite{TDR2, TDR4}.
The cross section for the $WW$-fusion process can also be determined by using the branching ratio for the $h \to b\overline{b}$ decay~\cite{TDR2}.
The obtained cross section values allow us to derive the branching ratio for the $h \to \tau ^+ \tau ^-$ decay.

At $\sqrt{s} = 250$ GeV, the contributions of the $WW$-fusion and $ZZ$-fusion  processes are negligible.
Therefore, we can use the Higgs-strahlung cross section to derive the branching ratio.
The Higgs-strahlung cross section $\sigma_{Zh}$ can be measured to a statistical precision of
$\Delta \sigma_{Zh} / \sigma_{Zh} = 2.5{\%}$ with the nominal TDR running scenario~\cite{TDR4}.
This improves to a subpercent level with the full running scenario~\cite{PhysicsCase}.

At $\sqrt{s} = 500$ GeV, both the Higgs-strahlung and the $WW$-fusion processes contribute to the Higgs boson production, whereas the contribution of the $ZZ$-fusion process is negligible.
For the $e^+e^-\to \nu \overline{\nu} h$ mode,
in which both processes are present,
it is in principle possible to estimate the contributions from the Higgs-strahlung and the $WW$-fusion processes separately, as discussed in Section 4.2.
However, we do not use this mode here for the estimate.
The expected statistical precision of the branching ratio after combining all the modes
except the $\nu \overline{\nu} h$ mode
is 3.6{\%} for the nominal running scenario.
This improves to 1.4{\%} with the full running scenario,
where we assume $\Delta \sigma_{Zh} / \sigma_{Zh} = 1.0{\%}$.

\begin{table*}[tb]
\centering
\caption{Expected precision of the cross section times the branching ratio
$\Delta (\sigma \times \mathrm{BR}) / (\sigma \times \mathrm{BR})$, assuming various running scenarios.
}
\begin{tabular}{cccccccc}\hline
Scenario & $\sqrt{s}$ (GeV) & $L$ (fb$^{-1}$) & $q \overline{q} h$ & $e^+ e^- h$ & $\mu ^+ \mu ^- h$ & $\nu \overline{\nu} h$ & Combined \\
\hline
Nominal \\
$\Delta (\sigma \times \mathrm{BR}) / (\sigma \times \mathrm{BR})$ & 250 & 250 & 3.4{\%} & 14.4{\%} & 11.3{\%} & --- & 3.2{\%} \\
 & 500 & 500 & 4.3{\%} & --- & --- & 6.9{\%} & --- \\
 & Combined & & 2.7{\%} & 14.4{\%} & 11.3{\%} & --- & 2.6{\%} \\
 & Combined & & --- & --- & --- & 6.9{\%} & 6.9{\%} \\
\hline
Initial \\
$\Delta (\sigma \times \mathrm{BR}) / (\sigma \times \mathrm{BR})$ & 250 & 500 & 2.5{\%} & 10.9{\%} & 8.7{\%} & --- & 2.4{\%} \\
 & 500 & 500 & 4.9{\%} & --- & --- & 9.6{\%} & --- \\
 & Combined & & 2.3{\%} & 10.9{\%} & 8.7{\%} & --- & 2.1{\%} \\
 & Combined & & --- & --- & --- & 9.6{\%} & 9.6{\%} \\
\hline
Full \\
$\Delta (\sigma \times \mathrm{BR}) / (\sigma \times \mathrm{BR})$ & 250 & 2000 & 1.3{\%} & 5.5{\%} & 4.3{\%} & --- & 1.2{\%} \\
 & 500 & 4000 & 1.7{\%} & --- & --- & 3.4{\%} & --- \\
 & combine & & 1.0{\%} & 5.5{\%} & 4.3{\%} & --- & 1.0{\%} \\
 & combine & & --- & --- & --- & 3.4{\%} & 3.4{\%} \\
\hline
\end{tabular}
\label{SummaryTable}
\end{table*}

\subsection{Systematic uncertainties}

The MC statistical uncertainties are found to have negligible impact on the results.
The systematic uncertainty in the luminosity measurement has been estimated to be 0.1{\%} or better for the ILC~\cite{lumi_error} and is not expected to be a significant source of systematic errors.
The uncertainties in the selection criteria, such as those caused by the uncertainty in the momentum/energy resolutions
and tracking efficiencies, are not included in this analysis, since they are beyond the scope of this paper.

\section{Summary}

We have evaluated the measurement precision of the Higgs boson production cross section times the branching ratio of decay into tau leptons at the ILC.
The study is based on the full detector simulation of the ILD model.
The dominant Higgs boson production mechanisms were studied at the center-of-mass energies of 250 GeV and 500 GeV, assuming the nominal luminosity scenario presented in the ILC TDR.
The analysis results are then scaled up to the running scenarios taking into account realistic running periods and a possible luminosity upgrade.

The results for the various modes and scenarios are summarized in Table~\ref{SummaryTable}.
In short, the cross section times the branching ratio can be measured with a statistical precision of $\Delta (\sigma \times \mathrm{BR}) / (\sigma \times \mathrm{BR}) = 2.6{\%}$ and $1.0{\%}$ for the nominal and full running scenarios, respectively.
We evaluate the statistical precision of $\mathrm{BR} (h \to \tau ^+ \tau ^-)$ to be 3.6{\%}
for the nominal TDR integrated luminosity and 1.4{\%} for the full running scenario, respectively.
These results serve to provide primary information
on the expected precision of measuring Higgs decays to tau leptons at the ILC,
which will be useful for future phenomenological studies on physics beyond the SM.

\begin{acknowledgements}
The authors would like to thank all the members of the ILC Physics Working Group.
We thank H. Yokoyama for providing the code for the signal event generator interfacing \verb|GRACE| and \verb|TAUOLA|.
This work has been partially supported by JSPS Grants-in-Aid for Science Research No. 22244031 and the JSPS Specially Promoted Research No. 23000002.
\end{acknowledgements}


\begin{thebibliography}{99}
\bibitem{ATLAS}
G. Aad \textit{et al.} [ATLAS Collaboration],
Phys. Lett. B \textbf{716} (2012) 1 - 29

\bibitem{CMS}
S. Chatrchyan \textit{et al.} [CMS Collaboration],
Phys. Lett. B \textbf{716} (2012) 30 - 61

\bibitem{deviate}
R. S. Gupta, H. Rzehak, J. D. Wells,
Phys. Rev. D \textbf{86} (2012) 095001

\bibitem{HTTevidence1}
G. Aad \textit{et al.} [The ATLAS Collaboration],
JHEP \textbf{04} (2015) 117

\bibitem{HTTevidence2}
S. Chatrchvan \textit{et al.} [The CMS Collaboration],
JHEP \textbf{05} (2014) 104

\bibitem{HTTevidence3}
The ATLAS and CMS Collaborations,
ATLAS-CONF-2015-044, CMS-PAS-HIG-15-002 (2015)

\bibitem{previous1}
J.-C. Brient, LC-PHSM-2002-003 (2002)

\bibitem{previous2}
M. Battaglia, arXiv:hep-ph/9910271 (1999)

\bibitem{TDR1}
T. Behnke \textit{et al.}, The International Linear Collider Technical Design Report Volume 1: Executive Summary (2013), arXiv:1306.6327 [physics.acc-ph]

\bibitem{TDR2}
H. Baer \textit{et al.}, The International Linear Collider Technical Design Report Volume 2: Physics (2013), arXiv:1306.6352 [hep-ph]

\bibitem{TDR3_1}
C. Adolphsen \textit{et al.}, The International Linear Collider Technical Design Report Volume 3.I: Accelerator R{\&}D in the Technical Design Phase (2013), arXiv:1306.6353 [physics.acc-ph]

\bibitem{TDR3_2}
C. Adolphsen \textit{et al.}, The International Linear Collider Technical Design Report Volume 3.II: Accelerator Baseline Design (2013), arXiv:1306.6328 [physics.acc-ph]

\bibitem{TDR4}
T. Behnke \textit{et al.}, The International Linear Collider Technical Design Report Volume 4: Detectors (2013), arXiv:1306.6329 [physics.ins-det]

\bibitem{PhysicsCase}
K. Fujii \textit{et al.}, arXiv:1506.05992 [hep-ex] (2015)

\bibitem{BeamPolOpe}
T. Barklow, J. Brau, K. Fujii, J. Gao, J. List, N. Walker, K. Yokoya, arXiv:1506.07830 [hep-ex] (2015)

%


\bibitem{Whizard}
W. Kilian, T. Ohl, J. Reuter, Eur. Phys. J. C \textbf{71} (2011) 1742

\bibitem{GRACE1}
F. Yuasa \textit{et al.}, Prog. Theor. Phys. Suppl. \textbf{138} (2000) 18 - 23 (arXiv:hep-ph/0007053)

\bibitem{GRACE2}
Minami-Tateya web page, \verb|http://www-sc.kek.jp|


\bibitem{GuineaPig}
D. Schulte, DESY-TESLA-97-08 (1997)

\bibitem{GDE}
\begin{verbatim}
http://ilc-edmsdirect.desy.de/ilc-edmsdirect/
item.jsp?edmsid=D00000000925325
\end{verbatim}

\bibitem{ISR}
M. Skrzypek, S. Jadach, Z. Phys. C \textbf{49} (1991) 577 - 584

\bibitem{Yokoyama}
H. Yokoyama, Master's thesis at the University of Tokyo (2014)

\bibitem{Tauola1}
S. Jadach, J. H. K\"{u}hn, Z. W\c{a}s, Comput. Phys. Commun. \textbf{64} (1991) 275 - 299

\bibitem{Tauola2}
P. Golonka, B. Kersevan, T. Pierzcha{\l}a, E. Richter-W\c{a}s, Z. W\c{a}s, M. Worek, Comput. Phys. Commun. \textbf{174} (2006) 818 - 835

\bibitem{Tauola3}
N. Davidson, G. Nanava, T. Przedzinski, E. Richter-W\c{a}s, Z. W\c{a}s, Comput. Phys. Commun. \textbf{183} (2012) 821 - 843

\bibitem{Pythia}
T. Sj\"{o}strand, S. Mrenna, P. Skands,
JHEP \textbf{0605} (2006) 026





\bibitem{overlay}
P. Chen, T. L. Barklow, M. E. Peskin, Phys. Rev. D \textbf{49} (1994) 3209 - 3227

\bibitem{PFA}
M. A. Thomson, Nucl. Instrum. Meth. A \textbf{611} (2009) 25 - 40

\bibitem{NNLO}
S. Dittmaier \textit{et al.}, [LHC Higgs Cross Section Working Group],
arXiv:1201.3084v1 [hep-ph] (2012)

\bibitem{Mokka}
P. Mora de Freitas, H. Videau,
LC-TOOL-2003-010 (2003)

\bibitem{Geant4}
S. Agostinelli \textit{et al.} [GEANT4 Collaboration],
Nucl. Instrum. Meth. A \textbf{506} (2003) 250 - 303

\bibitem{SGV}
M. Berggren,
arXiv:1203.0217 [physics.ins-det] (2012)

\bibitem{Marlin}
F. Gaede, Nucl. Instrum. Meth. A \textbf{559} (2006) 177 - 180

\bibitem{colapp}
R. K. Ellis, I. Hinchliffe, M. Soldate, J. J. Van Der Bij,
Nucl. Phys. B \textbf{297} (1988) 221 - 243

\bibitem{Durham}
S. Catani, Yu. L. Dokshitzer, M. Olsson, G. Turnock, B. R. Webber,
Phys. Lett. B \textbf{269} (1991) 432 - 438

\bibitem{TMVA}
P. Speckmayer, A. H\"{o}cker, J. Stelzer, H. Voss, J. Phys. Conf. Ser. \textbf{219} (2010) 032057

\bibitem{ROOT}
R. Brun, F. Rademakers, Proceedings AIHENP'96 Workshop, Lausanne, Sep. 1996, Nucl. Instrum. Meth. A \textbf{389} (1997) 81 - 86, \verb|http://root.cern.ch|

\bibitem{kT1}
S. Catani, Yu. L. Dokshitzer, M. H. Seymour, B. R. Webber,
Nucl. Phys. B \textbf{406} (1993) 187 - 224

\bibitem{kT2}
S. D. Ellis, D. E. Soper,
Phys. Rev. D. \textbf{48}, 3160 (1993)

\bibitem{FastJet}
M. Cacciari, G. P. Salam, G. Soyez,
arXiv:1111.6097v1 [hep-ph] (2011)

\bibitem{lumi_error}
D. M. Asner \textit{et al.}, arXiv:1310.0763 [hep-ph] (2013)
\end{thebibliography}
\end{document}